\documentclass{ametsocV5}

\usepackage{amsmath,amsfonts,amssymb,bm}
\usepackage{mathptmx}
\usepackage{newtxtext}
\usepackage{newtxmath}

\usepackage{color}
\definecolor{myred}{rgb}{0.9,0.0,0.1}

\usepackage[finalnew]{trackchanges} 

\title{Spectral wave energy dissipation due to under-ice turbulence}

\authors{Agnieszka Herman\correspondingauthor{Agnieszka Herman, oceagah@ug.edu.pl}}

\affiliation{Institute of Oceanography, University of Gda\'nsk, Poland\\ \ \\ \textcolor{myred}{This manuscript has been accepted at
Journal of Physical Oceanography.
Copyright in this work may be transferred without further notice.}}

\abstract{Dissipation within the turbulent boundary layer under sea ice is one of many processes contributing to wave energy attenuation in ice-covered seas. Although recent observations suggest that the contribution of that process to the total energy dissipation is significant, its parameterizations used in spectral wave models are based on fairly crude, heuristic approximations. In this paper, an improved source term for the under-ice turbulent dissipation is proposed, taking into account the spectral nature of that process (as opposed to parameterizations based on the so-called representative wave), as well as effects related to sea ice concentration and floe-size distribution, formulated on the basis of the earlier results of discrete-element modeling. The core of the new source term is based on an analogous model for dissipation due to bottom friction derived by Weber in 1991 (\url{https://doi.org/10.1017/S0022112091003634}). The shape of the wave energy attenuation curves and frequency-dependence of the attenuation coefficients are analyzed in detail for compact sea ice. The role of floe size in modifying the attenuation intensity and spectral distribution is illustrated by calibrating the model to observational data from a sudden sea ice break-up event in the marginal ice zone.
}

\begin{document}

\maketitle

%
%
%

%

\section{Introduction}\label{sec:intro}

As waves propagate through an ice-covered ocean, their energy is attenuated due to energy-conserving scattering and several dissipative processes, taking place within the ice itself and in the water column below it. Contrary to scattering, which has been extensively studied and can be regarded as well understood, the nature of dissipative processes remains relatively unexplored and modeling of their contribution to wave energy attenuation in different ice and forcing conditions remains a challenge. Existing models have problems with reproducing the observed rates and wave-frequency dependence of dissipation \citep{Meylanetal14,Meylanetal18,Squire18,Squire20,Shen19} and often require physically unrealistic values of coefficients to calibrate them to observational data \citep{Liuetal2020,Squire20}.

This paper concentrates on one of the arguably least explored wave energy dissipation mechanisms, namely the turbulent dissipation in the oscillatory boundary layer under the ice. Most observational and modelling studies of under-ice turbulence focus on the central ice pack or land-fast ice, when turbulence is related to the vertical shear of wind-induced currents, internal waves, tides, or buoyancy, i.e., spatial and temporal scales much larger than those associated with short-frequency, wind-generated  waves \citep[e.g.,][]{McPheeMartinson94,Skyllingstadetal09,Stevensetal09}. In fact, it is the absence of short waves -- a factor obstructing measurements, unavoidable in the open sea -- that makes the ice-covered ocean an attractive location for those studies \citep{McPheeMorison01}. Under-ice energy dissipation related to short waves was first considered by \cite{LiuMolloChristensen88}, who used a simple linear model to derive an attenuation term associated with viscous dissipation in water, assuming a constant viscosity coefficient, set to the kinematic viscosity of water. A crucial property of that model, resulting directly from its underlying assumptions, is an attenuation coefficient independent of wave amplitude, and thus an exponential form of the predicted attenuation curve. The model is unsuitable for turbulent dissipation. Nevertheless, owing to its simplicity, the solution by \cite{LiuMolloChristensen88} is used in spectral wave models, e.g. in Wavewatch~III, with the (low) kinematic viscosity replaced with (much higher) turbulent viscosity, often treated as a freely adjustable parameter \citep[e.g.,][]{RogersOrzech13,Ardhuinetal16}. Although this heuristic approach produces acceptable results, the lack of dependence of dissipation coefficient on wave amplitude explains some  difficulties with calibrating the models to both calm and storm conditions \citep{Lietal15}.

Limitations of this approach have been recognized e.g. by \cite{Stopaetal2016}, who computed the under-ice dissipation as a weighted average of laminar dissipation \citep[from the model of][]{LiuMolloChristensen88}, dominating at low Reynolds numbers, and turbulent dissipation, proportional to the amplitude of the orbital free-stream velocity under the ice and dominating at high Reynolds numbers. Their model has been used later by \cite{Boutinetal2018} in an analysis of the relative contribution of different physical mechanisms to modeled and observed wave attenuation in sea ice. The turbulent part of the model by \cite{Stopaetal2016} is based on an analogous formulation for the bottom boundary layer. The attenuation coefficient in those models depends on the total wave energy, which results in non-exponential attenuation curves. The same is true for the model by \cite{Kohoutetal11}, based on a simple quadratic drag law, and the discrete-element (DEM) models by \cite{Herman18} and \cite{Hermanetal19a,Hermanetal19b}. For monochromatic waves, those models predict the change of wave amplitude $a$ with distance $x$ as $\mathrm{d}a/\mathrm{d}x=-\alpha a^n$ with $n\neq1$, i.e., in the form analyzed recently by \cite{Squire18}.

The overall idea behind this paper is similar to that of \cite{Stopaetal2016}. The main goal is to develop a source term suitable for spectral wave models, describing wave energy dissipation within the oscillatory turbulent boundary layer under the ice and based on the existing, analogous source terms for dissipation by bottom friction. However, the formulation proposed here differs from the previous ones in two very important aspects. First, it is not based on the concept of a representative wave, underlying most bottom friction models \citep{Madsenetal88,Madsen94,Tolman1994,Zou2004} implemented in Wavewatch~III, SWAN (Simulating WAves Nearshore) and other spectral wave models, and also used in the algorithm by \cite{Stopaetal2016}. In the case of turbulent dissipation at the bottom, computing the attenuation coefficient from the `dominating', or representative wave properties -- as opposed to the whole frequency--direction spectrum -- is justified, because the velocity spectrum at the bottom is much narrower than at the surface \citep[only the long, low-frequency components reach the bottom;][]{Holthuijsen2007}. In sea ice, especially in regions not far from the ice edge, before the short waves are removed from the spectrum due to their strong dissipation, a more general approach is preferred, taking into account the shape of the wave energy spectrum. Such \add{a} model has been derived for bottom friction by \cite{Weber1991} and it is adopted here for the under-ice boundary layer. The second important aspect of the new formulation, mentioned above, is that it takes into account sea ice concentration and floe-size distribution. Obviously, dissipation within the turbulent boundary layer under the ice depends not on the oscillatory wave motion outside of that layer, but on the relative motion between ice and water, and the solutions for an (immovable) bed can be directly transferred to sea ice only when it is compact, confined horizontally, so that the amplitude of its horizontal motion is negligible. At ice concentrations allowing individual motion of ice floes, small floes follow the motion of the surrounding water and large floes remain almost stationary, making the ice--water friction strongly floe-size dependent. The wave-induced motion of ice floes of different sizes has been analyzed recently by \cite{Herman18} and \cite{Hermanetal19a,Hermanetal19b}, and their results are used here to formulate a `correction' for ice concentration and floe size to the basic dissipation source term, suitable for a compact ice cover.

The new source term is derived in the next section, which progresses from the description of the underlying assumptions through the presentation of the original model of \cite{Weber1991} to the formulation of dissipation under continuous ice and, finally,  fragmented ice with an arbitrary ice concentration. In Section~\ref{sec:atten}, the resulting wave energy attenuation is analyzed in detail, including the shape of the attenuation curves and frequency-dependence of the attenuation coefficients. The role of the floe-size distribution in modifying the intensity and spectral distribution of dissipation is illustrated by calibrating the model to observational data from a case study of \cite{Collinsetal15}. A discussion of the model features in the context of available observational data can be found in Section~\ref{sec:discuss}.

\section{Spectral dissipation due to boundary layer turbulence}\label{sec:model}

\subsection{Basic definitions and assumptions}

The source term formulated in parts \emph{c} and \emph{d} of this section is very general, suitable for implementation in spectral wave models for simulations with spatially-varying forcing, other source terms, etc. In this paper, however, it is tested in a highly simplified setting, as described below.

We consider random waves propagating through an ice cover extending in the $x$ direction from $x=0$ towards $x\rightarrow\infty$ and uniform in the $y$ direction, so that a one-dimensional energy transport equation can be solved, but the directionality of the energy spectra can be taken into account.

The ice cover is characterized by ice concentration $A_\mathrm{ice}$, area-weighted floe size distribution $\mathcal{P}_{\!\!a}(r_\mathrm{ice})$, where $r_\mathrm{ice}$ is floe radius ($\mathcal{P}_{\!\!a}$ is related to the more widely used number-weighted distribution $\mathcal{P}_{\!\!n}$ by $\mathcal{P}_{\!\!a}(r_\mathrm{ice}) = r_\mathrm{ice}^2\mathcal{P}_{\!\!n}(r_\mathrm{ice})/\int_0^\infty r_\mathrm{ice}^2\mathcal{P}_{\!\!n}(r_\mathrm{ice})\mathrm{d}r_\mathrm{ice}$), and the roughness length of the lower ice surface $z_0$.

The waves entering the ice at $x=0$ are linear, random-phase waves with energy spectrum $E_0(\theta,f)$, where $\theta$ denotes propagation direction \add{relative to the $x$ axis} and $f$ denotes frequency (with $\omega=2\pi f$ the angular frequency). In numerical simulations in this paper, $E_0$ is a JONSWAP spectrum with specified significant wave height $H_{s,0}$, peak period $T_{p,0}$, peak enhancement factor $\gamma_0$ and directional spreading $\sigma_{s,0}$ \citep{Holthuijsen2007}, or a multi-peaked combination of JONSWAP spectra. Discrete spectra are represented by $j=1,\dots,N_fN_d$ components, when $N_d$ is the number of directions uniformly spaced within the sector $[-\theta_\mathrm{lim},\theta_\mathrm{lim}]$ and $N_f$ is the number of frequencies logarithmically spaced between $f_\mathrm{min}$ and $f_\mathrm{max}$.

For each spectral component $j$ the stationary energy transport equation is:
\begin{equation}\label{eq:entrans}
  \frac{\mathrm{d}}{\mathrm{d}x_j}(c_{g,j}E_j) = S_{\mathrm{ice},j},
\end{equation}
where $x_j=x\cos\theta_j$ and $c_{g,j}$ denotes the group velocity of that component. Open-water dispersion relation is assumed, $\omega_j^2=gk_j\tanh[k_jh]$, so that $c_{g,j}=c_jn_j$ with $c_j=\omega_j/k_j$ the phase speed, $k_j$ the wavenumber and $n_j=\frac{1}{2}\left(1+2k_jh/\sinh[2k_jh]\right)$. The water depth $h$ is constant. The only source term considered, $S_{\mathrm{ice},j}$, represents turbulent dissipation in the under-ice boundary layer. It is assumed that it has the form:
\begin{equation}\label{eq:Sfice}
  S_{\mathrm{ice},j} = A_\mathrm{ice}S_{\mathrm{surf},j},
\end{equation}
i.e., dissipation takes place only in the ice-covered part of the domain. In sections \ref{sec:model}.\emph{\ref{sec:Ssurf}} and~\ref{sec:model}.\emph{\ref{sec:Ccr}} the term $S_{\mathrm{surf},j}$ is formulated for a continuous ice sheet with negligible wave-induced horizontal motion, and for an arbitrary ice concentration and floe size distribution, respectively. As the general form of $S_{\mathrm{surf},j}$ is based on $S_{\mathrm{bot},j}$ obtained by \cite{Weber1991}, the derivation is preceded in section~\ref{sec:model}.\emph{\ref{sec:Sbot}} by a concise description of his model.

\subsection{Spectral dissipation due to bottom friction}\label{sec:Sbot}

As mentioned in the introduction, $S_{\mathrm{surf},j}$ is formulated based on the eddy-viscosity bottom dissipation model by \cite{Weber1991}. The paper by Weber contains a very detailed derivation of the bottom dissipation source term $S_{\mathrm{bot},j}$. Here, only the final result is presented, together with the most important assumptions.

The essential part of the model is a formal parameterization of the turbulent stress, which is a generalization of simpler models based on a drag law and on eddy viscosity \citep{HasselmannCollins68,Madsenetal88,Madsen94}. The modifications of the flow within the water column, leading to energy dissipation, are expressed in terms of the (irrotational) zeroth-order flow at the top of the bottom boundary layer (known from the linear random wave theory) and the bottom stress, which has to be parameterized based on the zeroth-order solution. It is assumed that the boundary layer is fully turbulent, \change{its thickness is small compared to wavelength}{the ratio $\delta$ of its thickness to the wavelength is small, $\delta\ll1$}, and that the bottom surface is rough, so that the only relevant lengthscale characterizing the flow within the boundary layer is the equivalent Nikuradse roughness length $k_N$, related to the bottom roughness $z_0$ by $k_N=30z_0$. Within the boundary layer, the vertical variations in turbulent stresses are much larger than horizontal variations.

With those -- very general -- assumptions, the dissipation source term $S_{\mathrm{bot},j}$ can be obtained as a product of the velocity spectrum at the bottom and a proportionality factor $C_{\mathrm{bot},j}$ dependent on bottom stress parameterization:
\begin{equation}\label{eq:Sbot}
  S_{\mathrm{bot},j} = -C_{\mathrm{bot},j}\frac{\omega_j^2}{2g\sinh^2[k_jh]}E_j,
\end{equation}
In formulations used in spectral wave models, e.g. in SWAN, $C_{\mathrm{bot},j}=C_\mathrm{bot}$ for all $j$, i.e., the same value of the dissipation coefficient is used for all spectral components, and $C_\mathrm{bot}$ is either treated as an empirically adjustable constant or it is computed based on so-called representative characteristics of the spectrum \citep{Madsenetal88,Madsen94,Zou2004}. As discussed in the introduction, this computationally effective approximation is acceptable, as the near-bottom velocity spectra tend to be very narrow. In the model of \cite{Weber1991}, $C_{\mathrm{bot},j}$ is \change{obtained as a product of two terms, a velocity scale}{a function of the friction velocity} $u^\ast_\mathrm{bot}$ and a complex transfer function $T^\ast_j$ between the free-stream velocity outside of the boundary layer and stress within the boundary layer:
\begin{equation}\label{eq:Cf_weber}
  C_{\mathrm{bot},j} = u^\ast_\mathrm{bot}\check{T}_j(\varepsilon_0),
\end{equation}
\add{where $\check{T}_j(\varepsilon_0)=(T^\ast_j(\varepsilon_0)+\overline{T^\ast_j}(\varepsilon_0))$, $\overline{T^\ast_j}$ is the complex conjugate of $T^\ast_j$, and $T^\ast_j$ is defined below.} To evaluate \change{those two terms}{the terms in~}(\ref{eq:Cf_weber}), one additional assumption is necessary. Following \cite{Weber1991}, an eddy-viscosity model can be used, with \add{the turbulent shear} stress $\mathbf{\tau}(z)$ proportional to the vertical gradient of velocity $\mathbf{u}$, $\mathbf{\tau}(z)=\epsilon\frac{\partial\mathbf{u}}{\partial z}$, and $\epsilon=\kappa u^\ast_\mathrm{bot} z$, where $\kappa=0.4$ is the von K\'{a}rm\'{a}n constant. Then, \add{the stress $\tau$ is jointly Gaussian and} $u^\ast_\mathrm{bot}$ characterizes the variance and directional spreading of the bottom velocity spectrum, $u^\ast_\mathrm{bot}=\sigma_{11,\mathrm{bot}}^{1/2} \tilde{F}(1-\sigma_{22,\mathrm{bot}}/\sigma_{11,\mathrm{bot}})$, where $\tilde{F}(x)=\sqrt{2}\Gamma^2(\frac{5}{4})F_\mathrm{hg}^2(-\frac{1}{4},\frac{1}{2},1,x)$, $\Gamma$ and $F_\mathrm{hg}$ are the Gamma and the hypergeometric functions, respectively, and:
\begin{equation}\label{eq:sigma}
  \sigma_{\alpha\beta,\mathrm{bot}} = \int_\theta\int_\omega\frac{k_\alpha k_\beta}{k^2}T^\ast(\varepsilon_0)\overline{T^\ast}(\varepsilon_0)\frac{\omega^2}{\sinh^2[kh]}E(\theta,\omega)d\theta d\omega,
\end{equation}
where $k_1=k\cos\theta_r$, $k_2=k\sin\theta_r$ and the angle $\theta_r$ is chosen such that $\sigma_{\alpha\beta,\mathrm{bot}}=0$ for $\alpha\neq\beta$ \citep[see Appendix~A1 in][for a proof that $\theta_r$ can always be found to fulfill that condition]{Weber1991}\note{text added}. Note that in this formulation, $u^\ast_\mathrm{bot}$ is isotropic.

In the case of the eddy-viscosity model, the complex transfer function $T^\ast_j(\varepsilon)$ between the free-stream velocity outside of the boundary layer and stress within the boundary layer is given by:
\begin{equation}\label{eq:Tstar}
  T^\ast_j(\varepsilon) = \frac{\kappa\varepsilon}{2}\frac{K_1(\varepsilon\exp[\pi i/4])} {K_0(\varepsilon_0\exp[\pi i/4])},
\end{equation}
$K_0$ and $K_1$ are the 0th and 1st order modified Bessel functions, respectively, $\varepsilon(z)=(4k_jz/\kappa)^{1/2}$ and $\varepsilon_0=\varepsilon(z_0)$.

Notably, the thickness of the boundary layer obtained as part of the solution is \change{$\delta=u^\ast k/\omega$}{$<u^\ast_\mathrm{bot}>/\omega$, where $<u^\ast_\mathrm{bot}>$ denotes the average of $u^\ast_\mathrm{bot}$. As already mentioned, this scale has to be small compared to the wavelength for the model to be valid, i.e., $\delta=<u^\ast_\mathrm{bot}>k/\omega\ll1$}.

The source term $S_{\mathrm{bot},j}$ for two example wave energy spectra is shown in Fig.~\ref{fig:Sfexamples}.

\subsection{Dissipation under continuous ice cover}\label{sec:Ssurf}

As described in the previous section, the assumptions underlying the model of \cite{Weber1991} are very general. It is reasonable to assume that the surface turbulent boundary layer under an ice cover (provided that the wave forcing is sufficiently strong for turbulence to occur) has analogous basic properties to the bottom boundary layer, and thus that the wave energy dissipation under ice can be computed from the free-stream flow characteristics at the boundary of that layer. Consequently, equations analogous to (\ref{eq:Sbot})--(\ref{eq:Tstar}), reformulated in terms of velocity spectra at the surface, should be suitable for under-ice dissipation. We have:
\begin{equation}\label{eq:Ssurf}
  S_{\mathrm{surf},j} = -C_{\mathrm{surf},j}C_{\mathrm{rA},j}^2\frac{\omega_j^2}{2g}E_j,
\end{equation}
with:
\begin{equation}\label{eq:Csurf}
  C_{\mathrm{surf},j} = u^\ast_\mathrm{surf}\check{T}_j(\varepsilon_0),
\end{equation}
\begin{equation}\label{eq:ustarsurf}
  u^\ast_\mathrm{surf}=\sigma_{11,\mathrm{surf}}^{1/2} \tilde{F}(1-\sigma_{22,\mathrm{surf}}/\sigma_{11,\mathrm{surf}}),
\end{equation}
\begin{equation}\label{eq:sigmasurf}
  \sigma_{\alpha\beta,\mathrm{surf}} = \int_\theta\int_\omega\frac{k_\alpha k_\beta}{k^2}T^\ast(\varepsilon_0)\overline{T^\ast}(\varepsilon_0)C_{\mathrm{rA},j}^2\omega^2E(\theta,\omega)d\theta d\omega,
\end{equation}
and $\check{T}_j(\varepsilon_0)$ computed as previously. Anticipating the derivation in the next section, the function $C_{\mathrm{rA},j}=C_{\mathrm{rA},j}(r_\mathrm{ice},A_\mathrm{ice})$ has been introduced into~(\ref{eq:Ssurf}) and~(\ref{eq:sigmasurf}), representing effects related to floe size and ice concentration. In compact ice, $C_{\mathrm{rA},j}=1$.

\begin{figure}[t]
  \noindent\includegraphics[width=25pc]{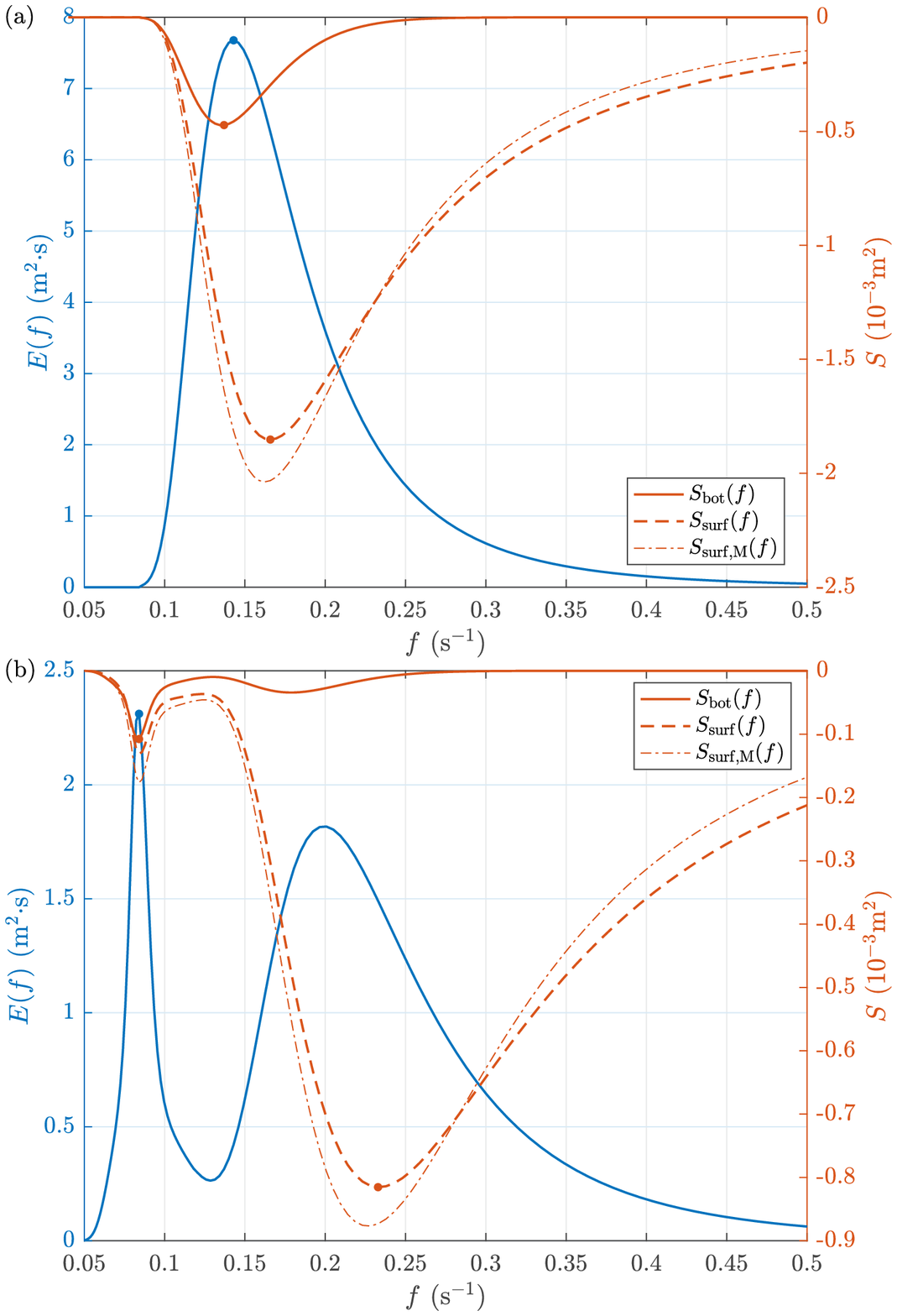}
  \caption{Dissipation source terms $S_{\mathrm{bot}}$ (continuous red lines) and $S_{\mathrm{surf}}$ (dashed red lines) for two example wave energy spectra (blue) in water depth $h=10$~m and $k_N=0.05$~m. For comparison, surface dissipation term computed with constant $C_\mathrm{surf}$\add{, leading to the same total dissipation,} is shown as well (marked as $S_{\mathrm{surf},\mathrm{M}}$).}\label{fig:Sfexamples}
\end{figure}

In Fig.~\ref{fig:Sfexamples}, $S_{\mathrm{bot},j}$ and $S_{\mathrm{surf},j}$ are compared for two example wave energy spectra. The two source terms have comparable amplitudes for the longest waves ($f<0.1$~s$^{-1}$ in the water depth considered), but, for obvious reasons, $S_{\mathrm{surf},j}$ has much larger values elsewhere in the spectrum and acts as a very effective low-pass filter, removing the short waves. Notably, maximum dissipation due to bottom/under-ice friction is shifted towards lower/higher frequencies relative to the peak of the spectrum, so that they contribute to the shift of the peak frequency in the opposite directions. In the case of $S_{\mathrm{surf},j}$, simple models with constant $C_\mathrm{surf}$ lead to underestimated/overestimated dissipation at high/low frequencies in comparison to the spectral formulation $C_{\mathrm{surf},j}$ (dashed and dotted-dashed lines in Fig.~\ref{fig:Sfexamples}; the effect is barely visible for $C_\mathrm{bot}$ and is therefore not shown).

\subsection{The correction for ice concentration and floe size}\label{sec:Ccr}

\begin{figure}[t]
  \noindent\includegraphics[width=35pc]{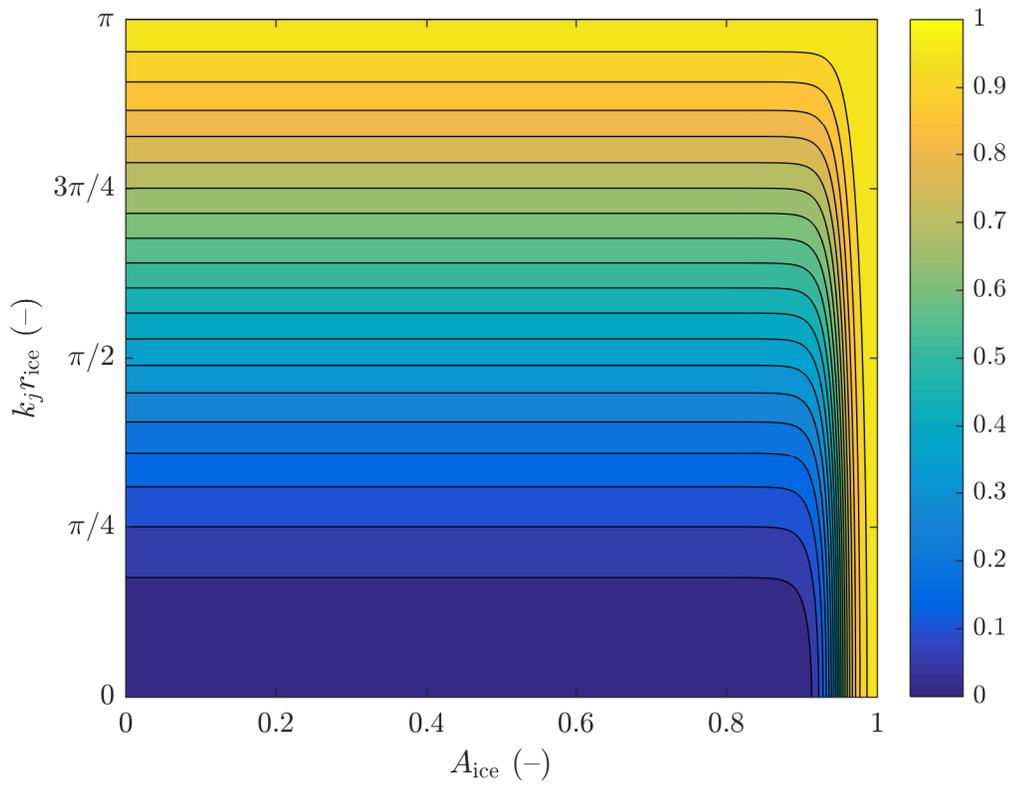}
  \caption{The term $C_{\mathrm{rA},j}$ computed from (\ref{eq:CrA}), (\ref{eq:zetaA}) with $A_\mathrm{lim}=0.95$, $\tilde{A}=0.01$ and $\mathcal{P}_{\!\!a}(r_\mathrm{ice})$ in the form of Dirac delta distribution.}\label{fig:CrA}
\end{figure}

Obviously, the formulation of $S_{\mathrm{surf},j}$ described in the previous section is acceptable only in compact, horizontally confined ice, in which the free-stream velocity at the boundary of the under-ice layer can be regarded as the relative ice--water velocity, which determines dissipation. The correction for ice concentration and floe size, proposed here, is based on the results of discrete-element simulations by \cite{Herman18} and \cite{Hermanetal19a,Hermanetal19b}, who analyzed patterns of wave-induced surge motion and collisions between ice floes of different sizes, as well as floe-size-dependent wave attenuation in the marginal ice zone (MIZ). The conclusions from those studies, relevant for the present discussion, are as follows. Wave-induced floe--floe collisions lead to strongly enhanced relative ice--water velocities and thus to increased dissipation due to bottom friction (depending on two main factors, the restitution coefficient of the ice and the effective ice--water friction coefficient). However, sustained collisions over larger areas require carefully adjusted, artificial conditions (confined model domain, spatially uniform wave forcing, etc.) and occur only within a narrow range of ice concentrations, separating the ``non-collisional regime'' at low $A_\mathrm{ice}$ (when ice floes move independently of each other) from the ``compact ice regime'' at very high $A_\mathrm{ice}$ (when ice floes stay in semi-permanent contact with their neighbors and the amplitude of their surge motion is very small). Moreover, in realistic settings with wave damping, collisions are limited to a narrow zone at the ice edge and play only a negligible role further down-wave. Crucially, for individual floes with diameter $2r_\mathrm{ice}$ forced by monochromatic waves with amplitude $a$ and wavenumber $k$, \cite{Herman18} showed that, in agreement with observations, the amplitude of their horizontal motion is $a\sin(kr_\mathrm{ice})/(kr_\mathrm{ice})$, and that this result is only weakly sensitive to the ice--water drag, i.e., in the equation of motion of the ice the inertial term is balanced by the force related to the wave-induced pressure, $\mathbf{F}_\mathrm{w}$. This finding can be easily extended to random waves under an assumption that the pressure induced by \change{indivudual}{individual} spectral components is additive. Then:
\begin{equation}\label{eq:movefloe}
  m_\mathrm{ice}\frac{\mathrm{d}\mathbf{u}_\mathrm{ice}}{\mathrm{d}t}=\sum_j\mathbf{F}_{\mathrm{w},j} = \sum_ja_j\omega_j^2m_\mathrm{ice}\frac{\sin[k_jr_\mathrm{ice}]}{k_jr_\mathrm{ice}} \frac{\mathbf{k}_j}{k_j} \sin\varphi_j,
\end{equation}
where $m_\mathrm{ice}$ is the mass of the floe, $\mathbf{u}_\mathrm{ice}$ its velocity and $\varphi_j$ denotes phase \citep[for derivation, see][]{Herman18}. Thus, the $j$th component of the relative ice--water velocity $\mathbf{u}_{\mathrm{r},j}$ is:
\begin{equation}\label{eq:urel}
  \mathbf{u}_{\mathrm{r},j} = \left(1-\frac{\sin[k_jr_\mathrm{ice}]}{k_jr_\mathrm{ice}}\right) \mathbf{u}_{\mathrm{surf},j},
\end{equation}
where $\mathbf{u}_{\mathrm{surf},j}$ is the free-stream surface velocity. This is valid for $A_\mathrm{ice}$ sufficiently low so that no contact between floes takes place. The transition from that ``sparse'' regime to the compact ice is very rapid and occurs at high ice concentration, generally $A_\mathrm{ice}>0.9$ \citep{Herman18}. Thus, the following expression is proposed for $C_{\mathrm{rA},j}$:
\begin{equation}\label{eq:CrA}
  C_{\mathrm{rA},j} = 1 - \zeta(A_\mathrm{ice}) \int_0^\infty\!\!\!\mathcal{P}_{\!\!a}(r_\mathrm{ice}) \frac{\sin[k_jr_\mathrm{ice}]}{k_jr_\mathrm{ice}}\mathrm{d}r_\mathrm{ice}
\end{equation}
with:
\begin{equation}\label{eq:zetaA}
  \zeta(A_\mathrm{ice}) = \frac{1}{2}\left(1-\tanh\left[\frac{A_\mathrm{ice}-A_\mathrm{lim}}{\tilde{A}}\right] \right)
\end{equation}
and $A_\mathrm{lim}$, $\tilde{A}$ adjustable coefficients. Figure~\ref{fig:CrA} shows $C_{\mathrm{rA},j}$ for $A_\mathrm{lim}=0.95$, $\tilde{A}=0.01$, computed with Dirac delta distributions as $\mathcal{P}_{\!\!a}(r_\mathrm{ice})$, i.e., constant $r_\mathrm{ice}$. As desired, $C_{\mathrm{rA},j}\rightarrow1$ for $A_\mathrm{ice}\rightarrow1$ independently of floe size, so that the solution for compact ice is recovered; and $\zeta(A_\mathrm{ice})\rightarrow1$ for $A_\mathrm{ice}\ll A_\mathrm{lim}$, recovering the size-dependent solution for sparsely distributed floes. Obviously, the form of~(\ref{eq:zetaA}) is arbitrary and another function with similar properties could be used as well. (The role of $\zeta(A_\mathrm{ice})$ is in many ways analogous to the role of the exponential term in the expression for internal ice pressure in sea ice rheology models: it provides a rapid transition from compact to freely drifting ice, but its exact form used in models does not result from physically-based arguments.)

\begin{figure}[t]
  \noindent\includegraphics[width=30pc]{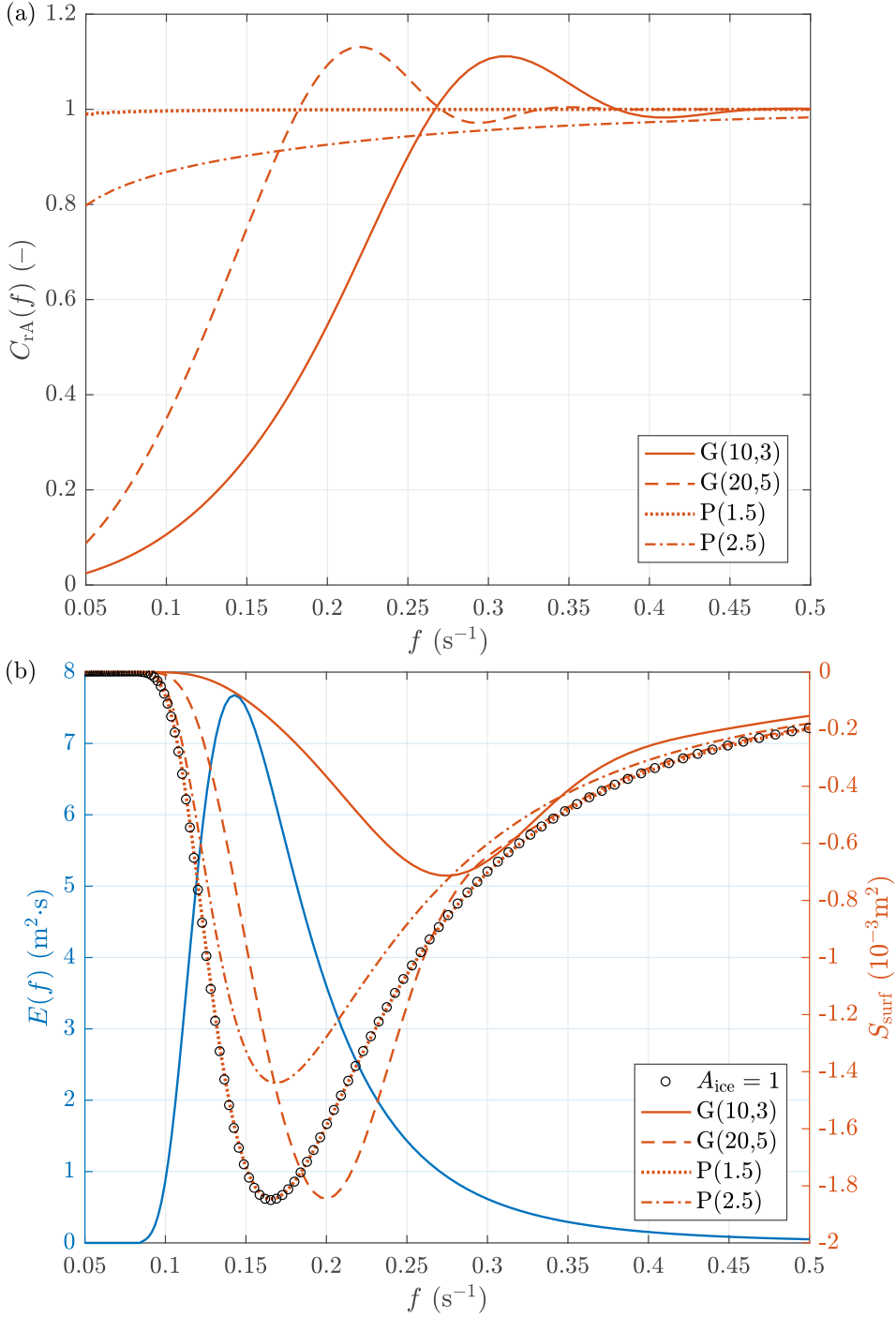}
  \caption{Function $C_\mathrm{rA}(f)$ for four different floe size distributions $\mathcal{P}_{\!\!a}$~(a) and the corresponding dissipation source term $S_{\mathrm{surf}}(f)$~(b) for the same spectrum as in Fig.~\ref{fig:Sfexamples}a. In the legend, `G' denotes Gaussian distribution with mean and standard deviation (in meters) in brackets, and `P' denotes power-law distribution with an exponent in brackets.}\label{fig:Sf_withCaR}
\end{figure}

Although the dependence of $C_\mathrm{rA}$ on wave frequency is very sensitive to the floe size distribution $\mathcal{P}_{\!\!a}$, only situations with relatively small floes lead to a drastic change of the resulting source term $S_\mathrm{surf}$ (Fig.~\ref{fig:Sf_withCaR}). When $\mathcal{P}_{\!\!a}$ is narrow (e.g., Gaussian) with a mean $\bar{r}_\mathrm{ice}$, dissipation of low-frequency waves with lengths larger than $\bar{r}_\mathrm{ice}$ is very weak. If the peak of the spectrum is located in that frequency range (as in the example in Fig.~\ref{fig:Sf_withCaR}), it is hardly affected, because the peak of dissipation is shifted towards higher frequencies. In short, small ice floes follow the motion of long waves, reducing the energy dissipation in low-frequency range, and can dissipate only the highest-frequency waves. This effect vanishes with increasing $\bar{r}_\mathrm{ice}$ -- in the analyzed example, with $\bar{r}_\mathrm{ice}=50$~m the dissipation source term $S_\mathrm{surf}$ is almost identical to that for $A_\mathrm{ice}=1$. When $\mathcal{P}_{\!\!a}$ widens, the frequency-dependence of $C_\mathrm{rA}$ becomes weaker. For a power-law $\mathcal{P}_{\!\!a}$, $C_\mathrm{rA}\rightarrow1$ as the exponent of the distribution decreases. As the exponent increases and the largest floes contribute less to the total sea ice area, their ability to dissipate the energy of the longest waves decreases as well (compare the dotted and dash-dotted lines in Fig.~\ref{fig:Sf_withCaR}).

\section{Wave energy attenuation due to under-ice turbulence}\label{sec:atten}

In this section, frequency-dependent attenuation rates are analyzed for a compact ice cover ($A_\mathrm{ice}=1$) and for loosely packed floes ($A_\mathrm{ice}\ll A_\mathrm{lim}$) with selected size distributions.

Assuming that the group velocity is constant in space and that $S_{\mathrm{ice},j}$ is computed from the set of equations~(\ref{eq:Ssurf})--(\ref{eq:sigmasurf}), (\ref{eq:CrA}) and (\ref{eq:zetaA}), the transport equation~(\ref{eq:entrans}) becomes:
\begin{equation}\label{eq:Eatten}
  \frac{1}{E_j}\frac{\mathrm{d}E_j}{\mathrm{d}x_j} = -A_\mathrm{ice}\frac{u^\ast_\mathrm{surf}}{g^2}C(k_jh)C_{\mathrm{aR},j}^2\check{T}_j(\varepsilon_0)\omega_j^3,
\end{equation}
where $C(k_jh)=2\cosh^2[k_jh]/(2k_jh+\sinh[2k_jh])$. In deep water, $C(k_jh)\rightarrow1$.

\subsection{Compact ice}

\begin{figure}[t]
  \noindent\includegraphics[width=30pc]{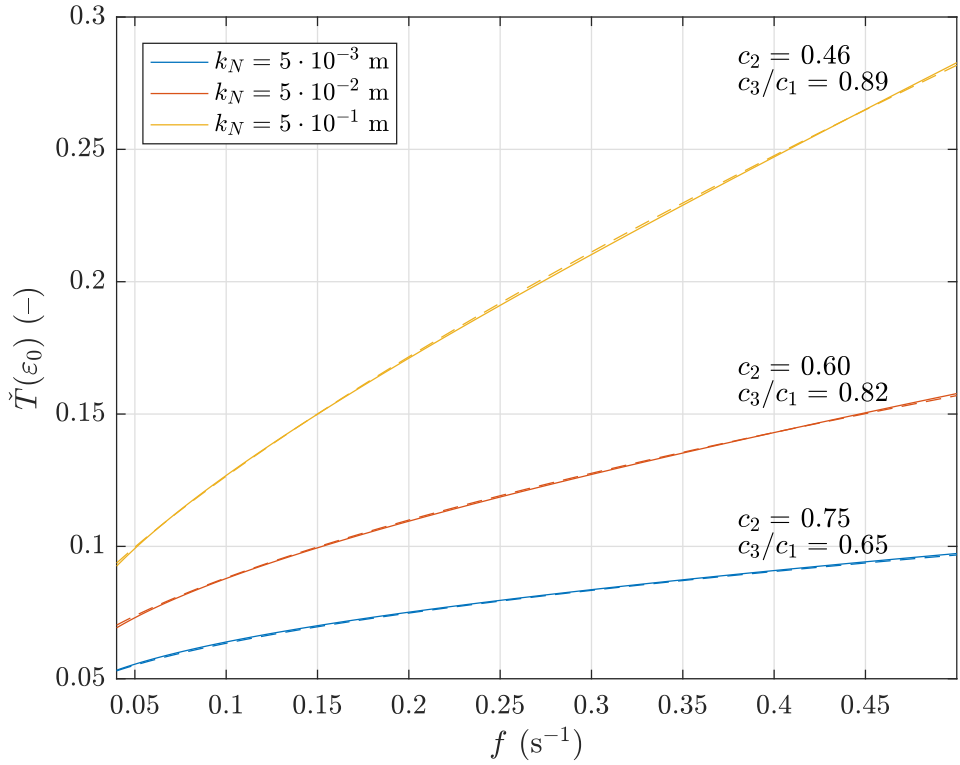}
  \caption{The term $\check{T}(\varepsilon_0)$ in function of wave frequency $f$ for three selected values of $k_N$ (colors). Dashed lines, barely visible under the continuous curves, show the least-square fit of function~(\ref{eq:Tstar_fit}) to $\check{T}(\varepsilon_0)$. The values of $c_2$ and $c_3/c_1$ accompanying the curves correspond to coefficients in~(\ref{eq:attenomega}).}\label{fig:Tstar}
\end{figure}

With $A_\mathrm{ice}=1$ and $C_{\mathrm{aR},j}=1$, the frequency-dependent part of the right-hand side of~(\ref{eq:Eatten}) is $\check{T}_j(\varepsilon_0)\omega_j^3$. Notably, apart from the wavenumber $k_j$, $\check{T}_j(\varepsilon_0)$ depends only on the Nikuradse roughness length $k_N$. It has been shown in Fig.~\ref{fig:Tstar} for three selected values of $k_N$ together with a least-square fit of a function:
\begin{equation}\label{eq:Tstar_fit}
  f_\mathrm{fit}(\omega,k_N) = (a_1k_N^{b_1})\omega^{a_2k_N^{b_2}} + a_3k_N^{b_3}.
\end{equation}
Thus, for a fixed $k_N$, $\check{T}_j(\varepsilon_0)$ can be approximated as $\check{T}_j(\varepsilon_0)\approx c_1\omega_j^{c_2}+c_3$, and the attenuation term in~(\ref{eq:Eatten}) is proportional to:
\begin{equation}\label{eq:attenomega}
  \omega_j^{3+c_2}+c_3/c_1\omega_j^3.
\end{equation}
For the three orders of magnitude of $k_N$ shown in Fig.~\ref{fig:Tstar}, $0.46\leq c_2\leq0.75$ and $0.65\leq c_3/c_1\leq0.89$. For instance, with $k_N=5\cdot10^{-2}$~m, which is a default value of this parameter in many spectral wave models, one has $\omega^{3.6}+0.82\omega^3$, i.e., the amplitude of both parts is comparable, but the first one is larger/smaller than the second one for deep-water wave periods below/above 8.7~s ($\omega=0.73$~rad$\cdot$s$^{-1}$), i.e., a slight increase of slope is observed between low and high frequencies.

Another important property of expression~(\ref{eq:Eatten}) is the fact that $u^\ast_\mathrm{surf}$ is not a constant, but a function of energy in all spectral components. For very narrow spectra, with energy concentrated within just a few frequency bins around the peak, $u^\ast_\mathrm{surf}\sim E_j^{1/2}$, i.e., the general form of equation~(\ref{eq:Eatten}) is $\mathrm{d}E_j/\mathrm{d}x_j=-\alpha_j(\omega)E_j^{3/2}$ rather than $\mathrm{d}E_j/\mathrm{d}x_j=-\alpha_j(\omega)E_j$. This is analogous to the solution obtained by \cite{Kohoutetal11} and \cite{Hermanetal19a} with a simple drag-law model (valid for monochromatic waves). For wide and/or multipeaked spectra we might expect $\mathrm{d}E_j/\mathrm{d}x_j=-\alpha_j(\omega)E_j^{n(\omega)}$, i.e., $n$ is frequency-dependent, with $n>1$ for all $\omega$. For frequencies around the dominating spectral peak, one might expect $n$ close to $3/2$. The expression for the energy attenuation $E_j(x_j)$ is:
\begin{equation}\label{eq:Eattenfinal}
  E_j(x_j)/E_{j,0} = \left[1+\tilde{\alpha}_jx\right]^{-1/(n_j-1)},
\end{equation}
with $\tilde{\alpha}_j=\alpha_j(n_j-1)E_{j,0}^{-(n_j-1)}$ (in m$^{-1}$) and $\alpha_j$ proportional to~(\ref{eq:attenomega}), but also dependent on the whole spectrum through $u_\mathrm{surf}^\ast$.

\begin{figure}[t]
  \noindent\ \ \includegraphics[width=28pc]{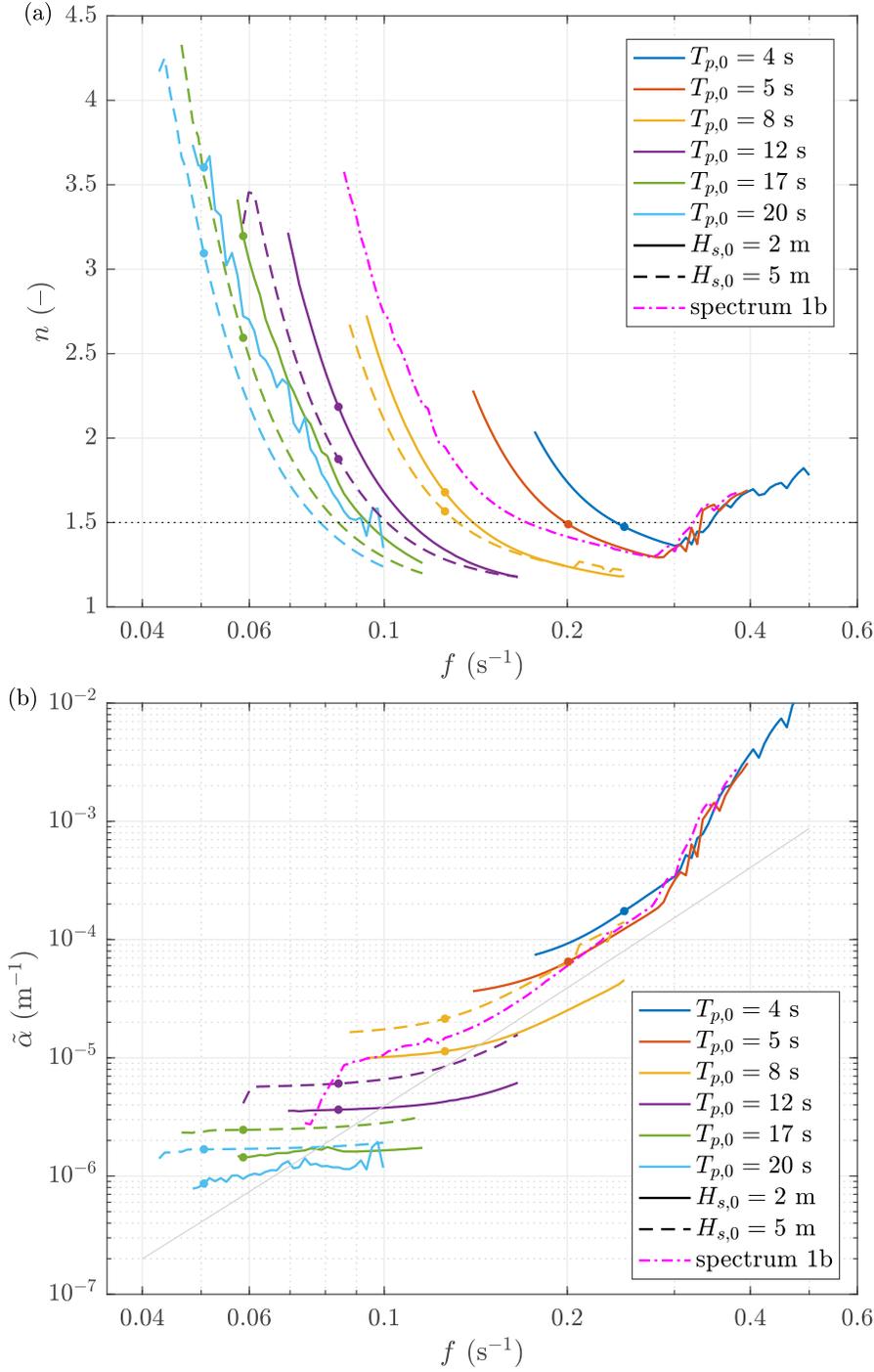}
  \caption{Least-square fitted values of $n$~(a) and $\tilde{\alpha}$~(b) in expression~(\ref{eq:Eattenfinal}) in function of wave frequency $f$ for a range of incident $T_{p,0}$ (colours) and $H_{s,0}$ (line styles). The colour dots mark the locations of the spectral peaks. Magenta lines correspond to the double-peaked spectrum in Fig.~\ref{fig:Sfexamples}b. Gray line in~(b) has slope given by expression~(\ref{eq:attenomega}). Results of simulations with compact ice.}\label{fig:n_alpha}
\end{figure}

The relationships described so far can be illustrated in more detail when equation~(\ref{eq:Eatten}) is solved numerically for a range of incident energy spectra with different $H_{s,0}$ and $T_{p,0}$. The resulting $n$ and $\tilde{\alpha}$ are shown in Fig.~\ref{fig:n_alpha}, and the example attenuation curves for an incident spectrum with $H_{s,0}=5$~m and $T_{p,0}=17$~s in Fig.~\ref{fig:Ex}. It is remarkable that the exponential function provides a particularly poor approximation of those curves around the spectral peak (blue line in Fig.~\ref{fig:Ex}), and thus also a poor approximation for $H_s(x)$. The approximation is inaccurate especially in the region close to the ice edge. As can be expected from the analysis above, the exponent $n$ generally decreases with frequency (Fig.~\ref{fig:n_alpha}a), so that the attenuation of the high-frequency part of any given spectrum is close to exponential: those components are attenuated very fast, but their influence on the total energy, and thus on $u_\mathrm{surf}^\ast$ is limited, so that they dissapear from the spectrum before $u_\mathrm{surf}^\ast$ substantially changes, making~(\ref{eq:Eatten}) for those components close to linear.

\subsection{Ice with $A_\mathrm{ice}<A_\mathrm{lim}$}

\begin{figure}[t]
  \noindent\includegraphics[width=30pc]{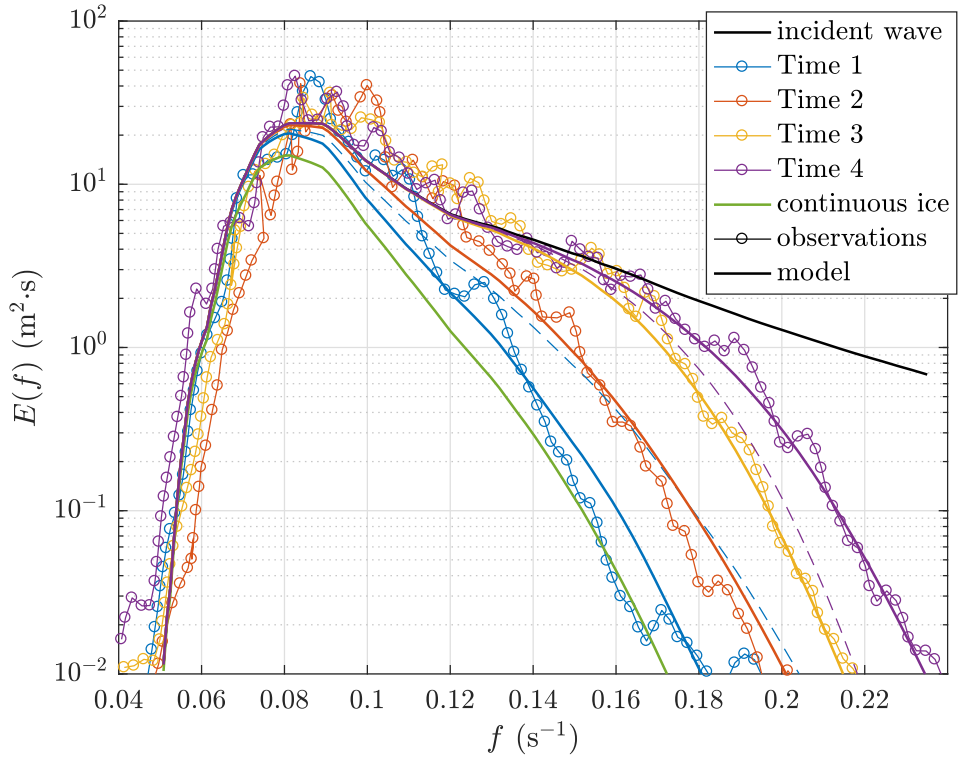}
  \caption{One-dimensional spectra $E(f)$ from the case study by \cite{Collinsetal15}. The black line shows the open water spetrum from SWAN, used as input to the model. Circles show measured spectra from four time instances (1: 2~May 23:24, 2: 3~May 00:30, 3: 3~May 02:30, 4: 3~May 03:30). Colour continuous lines show the corresponding results of the present model; results for continuous ice are shown in green. Two thin dashed lines show modelled spectra corresponding to Position~4 at Time~1 (blue) and Position~1 at Time~4 (violet); see text for details.}\label{fig:collins}
\end{figure}

As described in Section~\ref{sec:model}\emph{d}, in sea ice with
$A_\mathrm{ice}<A_\mathrm{lim}$ the floe-size distribution has a strong influence on the relative ice--water motion and thus on the under-ice wave energy dissipation. The data from the Barents Sea case study by \cite{Collinsetal15} will be used here to illustrate how the change in $\mathcal{P}_{\!\!a}$ modifies the penetration of storm waves into the ice cover. The analyzed event spans 4~hours between 23:24 on 2~May and 03:30 on 3~May 2010, when a sudden, storm-induced breakup of sea ice in the marginal ice zone led to a rapid increase of wave heights and a broadening of wave frequency spectra at the location of the ship R/V Lance, where the observations were made. During that time period, the ship moved gradually from the initial distance to the ice edge in the up-wave direction of $x_\mathrm{B}\approx100$~km to $x_\mathrm{E}\approx50$~km \citep[estimated from the map in Fig.~S5 of][]{Collinsetal15}. The open water wave energy spectrum from that event is available from the SWAN model, and it is used here as an input spectrum at $x=0$ (black line in Fig.~\ref{fig:collins}). Four measured spectra are available, labeled `Time~1'--`Time~4', and it is assumed that the ship location changed linearly from $x_\mathrm{B}$ at Time~1 to $x_\mathrm{E}$ at Time~4. Following the description by \cite{Collinsetal15}, it is assumed that the only factor that has changed during the analyzed four hours is the floe-size distribution. Therefore, the model is run several times with (arbitrarily) fixed $k_N=0.3$~m (corresponding to $z_0=0.01$~m, i.e., within the observed range for sea ice) and $A_\mathrm{ice}=0.9$, and with different $\mathcal{P}_{\!\!a}$ in order to find floe-size distributions with which the model optimally reproduces the observed wave energy spectra. \add{Two types of $\mathcal{P}_{\!\!a}$ are considered: upper-truncated power law  $\mathcal{P}_{\!\!a}\sim r_\mathrm{ice}^{-n_r}$ with the exponent $n_r=1.5$ and the maximum floe size $r_\mathrm{max}$ treated as an adjustable parameter, and a Gaussian distribution with adjustable mean $\mu_r$ and standard deviation $\sigma_r$. For each of the four stages of the event, the floe-size distribution is selected for which the simulated wave energy spectrum is closest to the observed one.}
The \change{simulated}{``optimal''} spectra \add{found in that way are} shown in Fig.~\ref{fig:collins}. \change{The corresponding floe-size distributions are}{They were obtained with}: $\mathcal{P}_{\!\!a}(r_\mathrm{ice})\sim r_\mathrm{ice}^{-1.5}$ with maximum floe radius of 150~m (Time~1); $\mathcal{P}_{\!\!a}(r_\mathrm{ice})\sim r_\mathrm{ice}^{-1.5}$ with maximum floe radius of 35~m (Time~2); Gaussian $\mathcal{P}_{\!\!a}(r_\mathrm{ice})$ with mean and standard deviation of 16~m and 5~m, respectively (Time~3); Gaussian $\mathcal{P}_{\!\!a}(r_\mathrm{ice})$ with mean and standard deviation of 12~m and 5~m, respectively (Time~4). For reference, Fig.~\ref{fig:collins} shows also the results for continuous ice (green line), which exhibit significant attenuation not only within the high-frequency part of the spectrum, but also around its peak. It is also worth stressing that the change of spectra recorded at the ship cannot be explained by changes of its location -- the difference between the modelled spectra at the innermost and outermost location is only a small fraction of the difference between observations (see the thin dashed lines in Fig.~\ref{fig:collins}).

Obviously, this procedure cannot be regarded as model validation, because no observed floe-size distributions are available. \change{(apart from photographs and an information that the typical floe size in the last phase of the event was 5--10~m)}{Importantly, however, the photographs and the qualitative information provided by }\cite{Collinsetal15} \add{clearly indicate that, first, large floes with sizes exceeding 100~m were dominating in the ice pack around the ship at the beginning of the event (i.e., they covered the majority of the surface area of the ice), and second, that the ice was ``clearly broken into smaller, more uniform, floes'' towards the end of the event, with typical floe sizes of 5--10~m, approximately one order of magnitude smaller than the peak wavelength}. \change{However, i}{I}t is \add{thus} clear that the model is able to reproduce the observed wave evolution during the analyzed case with realistic floe-size distributions, closely corresponding to the qualitative description in \cite{Collinsetal15} and in agreement with their interpretation of the event. \remove{This issue will be further discussed in Section~4.} \add{The four optimal floe-size distributions found by minimizing the differences between the modeled and observed wave energy spectra progress from wide, power-law $\mathcal{P}_{\!\!a}$s towards narrow $\mathcal{P}_{\!\!a}$s and, importantly, towards smaller and smaller floes. In the model, it is the gradual removal of the largest floes that is crucial for reproducing the observed spectra and for shifting the dissipation towards higher and higher frequencies. (In fact, the detailed shape of $\mathcal{P}_{\!\!a}$ in the limit of very small floes is not relevant in this case, as they do not contribute to attenuation due to their small $k_jr_\mathrm{ice}$ ratio. Thus, more complex shapes of $\mathcal{P}_{\!\!a}$, e.g., those observed in laboratory ice broken by waves by} \cite{Hermanetal18}\add{, could be used equally well. In this study, Gaussian $\mathcal{P}_{\!\!a}$s are used for the sake of simplicity.) Finally, it is worth stressing that the results obtained are in agreement with the general knowledge of the observed variability of the floe-size distribution in sea ice and processes shaping it. Heavy-tailed, power law $\mathcal{P}_{\!\!a}$s with an exponent $n_r>2.0$ are widely observed in the marginal ice zone} \citep[see, e.g.,][and references there]{Herman10,Toyota16}. \add{Wave-induced breaking, on the other hand, tends to produce ice floes of similar sizes, dependent on the wave forcing and, predominantly, thickness and material properties of the ice} \citep{Squire95,Herman17}.

\section{Discussion}\label{sec:discuss}

The main aim of this paper was to formulate a source term for the wave energy transport equation, accounting for dissipation of wave energy within the turbulent boundary layer under sea ice. The formulation is based on an analogous solution for the bottom boundary layer derived by \cite{Weber1991}, with corrections for ice concentration and floe size distribution. Crucially, the new source term can be easily implemented in spectral wave models or in coupled wave--ice models. The only required information on sea ice properties are: the ice concentration, floe size distribution (or its estimate, e.g., the representative, or dominating floe size) and a measure of roughness of the lower ice surface. Arguably, the third one of those three variables is particularly hard to estimate, especially that it likely exhibits strong spatial variability. In practical applications of the source term, the roughness length is a natural candidate for an adjustable coefficient, determined by calibrating the model to observational data -- similarly as, e.g., the viscoelastic properties of the ice in the study of \cite{Chengetal17b} or \cite{Liuetal2020}. Although that kind of model calibration and validation remains to be done, it is worth stressing that, as the theoretical analysis above has shown, physically realistic values of the roughness length, within the range used in models of the bottom dissipation, produce realistic values of wave energy attenuation, in the order of $10^{-6}$--$10^{-5}$~m$^{-1}$ for long waves, with periods larger than 10~s, and $10^{-4}$--$10^{-3}$~m$^{-1}$ for waves with periods of a few seconds. The model could be also successfully adjusted to reproduce the observed evolution of wave energy spectra in the case study discussed in Section~\ref{sec:atten}\emph{b}. It is tempting to conclude that the model -- and thus the under-ice turbulent dissipation -- explains the observed variability of wave attenuation in that case. However, it is in fact very unlikely that a single process is responsible for energy dissipation in any realistic situation, and a successful calibration of any model taking into account only one process is a sign of our ignorance regarding ``true'', or realistic values of its parameters. The problem has been recently described by \cite{Hermanetal19b}, who found out that several different combinations of model parameters were comparably successful in reproducing wave attenuation patterns observed in a laboratory. In the present case, when the properties of the floe-size distribution are allowed to vary together with the ice roughness $z_0$ (which was fixed in the computations in Section~\ref{sec:atten}\emph{b}), it is likely that the model can be fitted to a very wide range of situations. When not one, but several dissipative source terms are included in a wave model, the number of unknown parameters and their possible combinations increases considerably, making any inferences about the relative importance of those source terms problematic. The general conclusion is that concurrent measurements of many different variables, not only of wave energy spectra, are essential for that type of analysis.

As far as under-ice turbulence is concerned, although its relative contribution to the total wave energy dissipation remains difficult to quantify, it is reasonable to assume that that contribution is substantial. Dissipation due to bottom friction is regarded as the dominant mechanism of dissipation in shallow shelf seas, and the corresponding source term is indispensable in coastal wave models \citep{Holthuijsen2007} -- it is thus unjustified to assume that, given orbital velocities under the ice comparable or even higher than those at the bottom, the under-ice friction is negligible. Apart from this very general argument, there is growing observational evidence for the role of turbulence in wave attenuation in sea ice, see, e.g., \cite{Voermansetal19} and \cite{SmithThomson19b}. \citep[An argument to the contrary has been presented in][but the contradictory interpretations have been reconciled in the later study.]{SmithThomson19} Similarly, \cite{Boutinetal2018}, who analyzed numerically the case study of \cite{Collinsetal15}, found that basal friction is indispensable to explain attenuation patterns observed during that event. Notably, they speak of ``nonlinear dissipation that vanishes when the ice is broken'', but they incorrectly treat inelastic dissipation within sea ice as the only nonlinear dissipative process, not recognizing that even the relatively simple friction model they use is nonlinear as well. Indeed, the attenuation coefficient in the model of \cite{Stopaetal2016}, used by \cite{Boutinetal2018}, depends on wave energy through its dependence on the amplitude of orbital velocity, making the resulting energy attenuation nonexponential. In fact, no reasonable model of turbulent friction is linear \citep[unlike the viscous friction suitable for laminar flows;][]{LiuMolloChristensen88}.

\begin{figure}[t]
  \noindent\includegraphics[width=30pc]{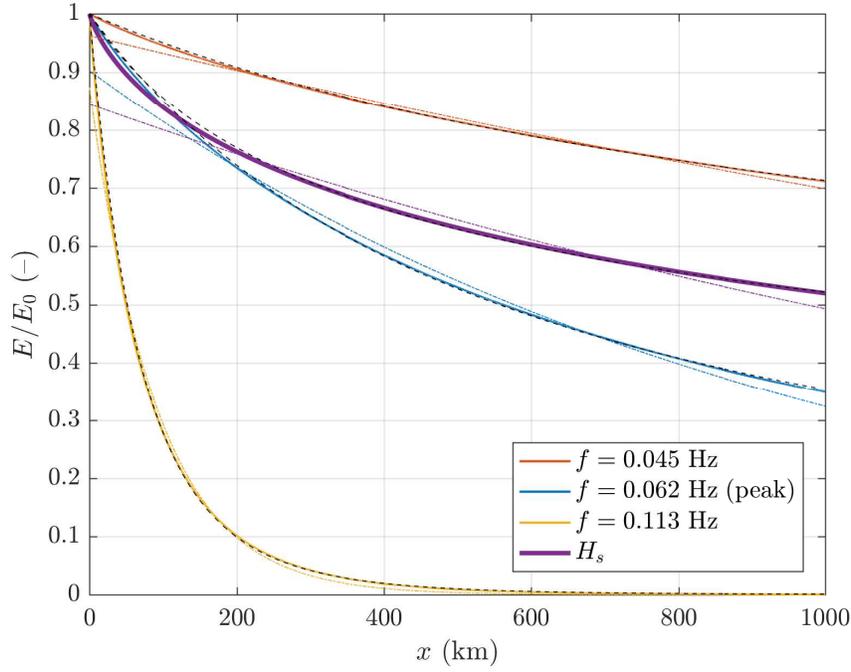}
  \caption{Attenuation curves (continuous lines) for an incident JONSWAP spectrum with $H_{s,0}=5$~m and $T_{p,0}=17$~s, for the energy at three selected frequencies and for the significant wave height. Dashed-dotted lines show the corresponding exponential fit, and black dashed lines the fit of expression~(\ref{eq:Eattenfinal}). \add{Both fits are computed over the whole range of $x$ shown in the plot}. Results of simulations with compact ice.}\label{fig:Ex}
\end{figure}

The question of the form of wave attenuation curves in different conditions -- beside the frequency-dependence of attenuation coefficients -- is an issue increasingly often discussed and investigated theoretically and numerically \citep{Squire18,Hermanetal19a,Hermanetal19b}, but very hard to resolve based on existing observational data. As concurrent observations are usually available at a limited number of locations, exponential attenuation curves are assumed \emph{a priori}, and the attenuation rates are computed either by least-square fitting an exponential function to data or, when only two data points are available, by computing an apparent attenuation $\alpha_\mathrm{ap}=\log(E_2/E_1)/(x_1-x_2)$ \citep[see, e.g.,][]{Meylanetal14,Rogersetal2016,Stopaetal18}. The attenuation curves predicted by the models of turbulent ice--water friction are steep close to the ice edge and much less steep further down-wave (Fig.~\ref{fig:Ex}), but recording a similar pattern in the field would require densely placed sensors, especially in the outer regions of the MIZ. Notably, when exponential curves are fitted to the numerical results obtained with the present model, the slopes within the low-frequency range $f<0.3$~Hz (typically available from observations) varies with frequency as $\omega^{-3.2}$--$\omega^{-3.4}$, which is within \add{the range} of observations.

Finally, it is worth noting that the model of \cite{Weber1991}, on which the present work is based, is formulated in a very general way and the eddy-viscosity parameterization used here is just a one special case out of several possible formulations. This makes the presented source term easily modifiable, e.g., when observational data become available supporting another type of parameterization of under-ice turbulence. As far as the spectral (as opposed to monochromatic) turbulent dissipation is concerned, it might be particularly important when considered in combination with other processes (scattering, nonlinear wave--wave interactions, other dissipation mechanisms) which are sensitive to the shape of the spectrum. Obviously, this type of analysis requires implementation of the present source term in a spectral wave model.

\acknowledgments
This work has been financed by Polish National Science Centre project no. 2018/31/B/ST10/00195 (``Observations and modeling of sea ice interactions with the atmospheric and oceanic boundary layers'').

%
%
\datastatement
The Matlab scripts used to generate the results presented in this paper can be obtained from the author.

%






%
%
%
\bibliographystyle{ametsoc2014}
\bibliography{SeaIce,Waves}

%

%

\end{document}